\documentclass{llncs}
\usepackage{amsfonts}
\usepackage[dvips]{graphicx}
\usepackage{amsmath}
\usepackage{amssymb}
\usepackage{xspace}
\usepackage{colortbl}

\title{Efficient Pattern Matching on Binary Strings}

\author{Simone Faro\inst{1} \and Thierry Lecroq\inst{2}}
\institute{Dipartimento di Matematica e Informatica, Universit\`a di Catania, Italy\\
\and
University of Rouen, LITIS EA 4108, 76821 Mont-Saint-Aignan Cedex, France\\
\email{faro@dmi.unict.it, thierry.lecroq@univ-rouen.fr}
}

\newlength{\defbaselineskip}
\newcommand{\setlinespacing}[1]%
           {\setlength{\baselineskip}{#1 \defbaselineskip}}

\newlength{\mylength}

\newcommand{\tb}[1]{\textbf{#1}}

\newcommand{\bigO}{\mathcal{O}}

\newcommand{\bm}{\textsc{Boyer-Moore}\xspace}
\newcommand{\bbm}{\textsc{Binary-Boyer-Moore}\xspace}
\newcommand{\naive}{\textsc{Binary-Naive}\xspace}
\newcommand{\bsks}{\textsc{Binary-Skip-Search}\xspace}
\newcommand{\bh}{\textsc{Binary-Hash-Matching}\xspace}
\newcommand{\ppp}{\textsc{Preprocess}\xspace}
\newcommand{\qh}{\textsc{$q$-Hash}\xspace}
\newcommand{\sks}{\textsc{Skip-Search}\xspace}

\newcommand{\ul}[1]{\underline{#1}}

\def\Hs{\mathit{Hs}}
\def\shift{\mathit{shift}}
\def\Mask{\mathit{Mask}}
\def\Patt{\mathit{Patt}}
\def\Last{\mathit{Last}}

\begin{document}

\maketitle

\begin{abstract}
The \emph{binary string matching} problem consists in finding all the occurrences of a pattern in a text where  both strings are built on a binary alphabet. This is an interesting problem in computer science, since binary data are  omnipresent in telecom and computer network applications. Moreover the problem finds applications also in the field of image processing and in pattern matching on  compressed texts. Recently it has been shown that adaptations of classical exact string matching algorithms are not very efficient  on binary data.
In this paper we present two efficient algorithms for the problem adapted to completely avoid any reference to bits allowing to process pattern and text byte by byte. Experimental results show that the new algorithms outperform existing solutions in most cases.\\

\end{abstract}

\section{Introduction}

Given a text $t$ and a pattern $p$ over some alphabet $\Sigma$ of size $\sigma$,  the \emph{string matching problem} consists in finding \emph{all} occurrences  of the pattern $p$ in the text $t$. It is a very extensively studied problem in computer science, mainly due to its  direct applications to such diverse areas as text, image and signal processing,  speech analysis and recognition, information retrieval, computational biology  and chemistry, etc.
%Several string matching algorithms have been proposed over the years~\cite{CL2004}, however the Boyer-Moore algorithm~\cite{BM77} deserves a special mention, since it has  been particularly successful and has inspired much work.

In this article we consider the problem of searching for a pattern $p$ of  length $m$ in a text $t$ of length $n$, with  both  strings are built over a binary alphabet, where each character  of $p$ and $t$ is represented by a single bit. Thus memory space needed to represent $t$ and $p$ is, respectively, $\lceil n/8 \rceil$  and $\lceil m/8 \rceil$ bytes.

This is an interesting problem in computer science, since binary data are  omnipresent in telecom and computer network applications. Many formats for data exchange between  nodes in distributed computer systems as well as most network protocols  use binary representations. Binary images often arise in digital image processing as masks or as the results of certain operations such as segmentation, thresholding and dithering. Moreover some input/output devices, such as laser printers and fax machines, can only handle binary images.

The main reason for using binaries is size. A binary is a much more compact format than the symbolic or textual  representation of the same information. Consequently, less resources are required to transmit binaries  over the network. For this reason the binary string matching problem finds applications also in pattern matching on  compressed texts, when using the Huffman compression strategy~\cite{KS2005,SD06,FG06}.

Observe that the text $t$, and the pattern $p$ to search for, cannot be  directly processed as strings with a super-alphabet~\cite{Fre2002,FG06},  i.e., where each group of 8 bits  is considered as a character of the text. This is because an occurrence of the pattern can be found starting at the middle
 of a group. Suppose, for instance, that $t =$ \texttt{011001001000100110100101000101001001} and  $p =$ \texttt{0100110100}. If we write text and pattern as groups of 8 bits then we obtain

\begin{center}
\begin{footnotesize}
\begin{tabular}{ll}
$t=$ & \texttt{01100100 10001001 10100101 00010100 1001}\\
$p=$ & \texttt{\ \ \ \ \ \ \ \ \ \ \ \ 01001 10100}\\
\end{tabular}
\end{footnotesize}
\end{center}

The occurrence of the pattern at position 11 of the text cannot be located by a classical pattern matching algorithm based on super-alphabet.

It is possible to simply adapt classical efficient algorithms for exact  pattern-matching to binary-matching with minor modifications. We can substitute in the algorithms reference to the character  at position $i$ with reference to the bit at position $i$. Roughly speaking we can substitute occurrences  of $t[i]$ with \textsc{getBit}$(t,i)$ which returns the $i$-th bit of the text $t$. This transformation does not affect the time complexity of the algorithm but  is time consuming and in general could be not very efficient.

In~\cite{KBN2007} Klein and Ben-Nissan proposed  an efficient variant of the \bm algorithm for the binary case without referring to bits. The algorithm is projected to process only entire blocks such as bytes or words and achieves a significantly reduction in the number of text character inspections. In~\cite{KBN2007} the authors showed also by empirical results that the new variant performs better than the regular binary \bm algorithm and even than binary versions of the most effective algorithms for classical pattern matching.

%how, by means of  some pre­computed tables, one may
%For instance, q-Hashed algorithms~\cite{Lec07} can be implemented efficiently by taking the entire sequence of $q$ bits as hash value. This permits to make no test to be sure that an occurrence has been found. As another example, we can naturally adapt bit-parallel solutions to binary search. Moreover we can also use solutions based on super-alphabets if such solutions take in to account also occurrences beginning at any bit position.

%In this paper we present results along those three lines: adaptation of classical string matching algorithms, adaptation of bit-parallel string matching algorithms and adaptation of string matching algorithms based on super alphabet to deal with binary data. We exhibit a very efficient solution using a super alphabet. It consists in grouping the bits of the pattern into groups of 8 bits  starting from each position between 0 and 7. We obtain thus 8 new patterns built on a 256-letter alphabet. Each pattern has a head and a tail that has within 0 and 7 bits. The 8 new patterns are searched for simultaneously in the text (accessed byte  per byte thus considered as built on a 256-letter alphabet)  using an efficient multiple string matching algorithms. Experimental results show that this new solution outperforms  existing solutions including recent ones (see \cite{KBN2007}) in most cases.

In this note we present two new efficient algorithms for matching on binary strings which, despite their $\bigO(nm)$ worst case time complexity, obtain very good results in practical cases. The first algorithm is an adaptation of the \qh algorithm~\cite{Lec07} which is among the most efficient algorithms for the standard pattern matching problem. We show how the technique adopted by the algorithm can be naturally translated to allow for blocks of bits.

The second solution can be seen as an adaptation to binary string matching of the \sks algorithm~\cite{CLP98}. This algorithm can be efficiently adapted to completely avoid any reference to bits allowing to process pattern and text proceeding byte by byte.

The paper is organized as follows. In Section~\ref{sec:definitions} we introduce basic definitions and the  terminology used along the article.
In Section~\ref{sec:high_level} we introduce a high level model used to process binary strings avoiding any reference to bits. Next, in Section~\ref{sec:newVariant}, we introduce the new solutions. Experimental data obtained by running under various conditions all the  algorithms reviewed are presented and compared in Section~\ref{sec:exp}. Finally, we draw our conclusions in Section~\ref{sec:conclusion}.

\section{Preliminaries and basic definitions}\label{sec:definitions}
A string $p$ of length $m\ge 0$ is represented as a finite array $p[0\,..\,m-1]$ of characters from a finite alphabet $\Sigma$.  In particular, for $m = 0$ we obtain the empty string, also denoted by $\varepsilon$.  By $p[i]$ we denote the $(i+1)$-th character of $p$, for $0\leq i < m$. Likewise, by $p[i\,..\,j]$ we denote the substring of $p$ contained between the $(i+1)$-th and the $(j+1)$-st characters of $p$, for $0\leq i \leq j < m$.  Moreover, for any $i,j \in \mathbb{Z}$, we put $p[i\,..\,j] = \varepsilon$ if $i>j$ and $p[i\,..\,j] = p[\max(i,0),\min(j,m-1)]$ if $i\leq j$. A substring of $p$ is also called a \emph{factor} of $p$.
A substring of the form $p[0\,..\,i]$ is called a \emph{prefix} of $p$ and a substring of the form $p[i\,..\,m-1]$ is
 called a \emph{suffix} of $p$ for $0\le i \le m-1$.
For any two strings $u$ and $w$, we write $w \sqsupseteq u$ to indicate that $w$ is a suffix of $u$.  Similarly, we write $w \sqsubseteq u$ to indicate that $w$ is a prefix of $u$.

Let $t$ be a text of length $n$ and let $p$ be a pattern of length $m$.  When the character $p[0]$ is aligned with the character $t[s]$ of the text, so that the character $p[i]$ is aligned with the character $t[s+i]$, for $i=0,\ldots,m-1$, we say that the pattern $p$ has \emph{shift} $s$ in $t$.  In this case the substring $t[s\,..\,s+m-1]$ is called the \emph{current window} of the text.  If $t[s\,..\,s+m-1] = p$, we say that the shift $s$ is \emph{valid}. %Thus the string matching problem can be rephrased as the problem of finding \emph{all} valid shifts of a pattern $p$ relative to a text $t$.

Most string matching algorithms have the following general structure. First, during a \emph{preprocessing phase}, they calculate useful mappings,
generally in the form of tables, which later are accessed to determine nontrivial shift advancements.  Next, starting with shift $s=0$, they look for all valid shifts, by executing a \emph{matching phase}, which determines whether the shift $s$ is valid and computes a \emph{positive} shift increment,
$\Delta s$. Such increment $\Delta s$ is used to produce the new shift $s+\Delta s$ to be fed to the subsequent matching phase.
%Observe that for the correctness of the algorithm it is plainly necessary that each shift increment $\Delta s$ computed is \emph{safe}, namely the interval $\{s+1,\ldots,s+\Delta s - 1\}$ contains no valid shifts.

For instance, in the case of the \textsc{Naive} string matching algorithm, there is no preprocessing phase and the matching phase always returns a unitary shift increment, i.e. all possible shifts are actually processed.

%In contrast the \bm algorithm~\cite{BM77} checks whether $s$ is a valid shift, by scanning the pattern $p$ from right to left and, at the end of the matching phase, it computes the shift increment as the maximum value suggested by two heuristics: the \emph{good-suffix heuristic} and the \emph{bad-character heuristic}, provided that both of them are applicable (see~\cite{CL2004}).

\section{A High Level Model for Matching on Binary Strings}\label{sec:high_level}
A string $p$ over the binary alphabet $\Sigma=\{0,1\}$ is said to be a \emph{binary string} and is represented as a binary vector $p[0\,..\,m-1]$, whose elements are bits. Binary vectors are usually structured in blocks of $k$ bits, typically bytes ($k=8$), half­words ($k=16$) or words ($k=32$), which can be processed at the cost of a single operation. If $p$ is a binary string of length $m$ we use the symbol $P[i]$ to indicate the $(i+1)$-th block of $p$ and use $p[i]$ to indicate the $(i+1)$-th bit of $p$. If $B$ is a block of $k$ bits we indicate with symbol $B_j$ the $j$-th bit of $B$, with $0\leq j< k$. Thus, for $i=0,\ldots,m-1$ we have $p[i]=P[\lfloor i/k \rfloor]_{i\mod k}$.

%A method to solve the pattern matching problem on binary strings could consists in allowing standard pattern matching algorithms to process binary strings bit-by-bit by using bitwise operations. However the sequential access to single bits of a binary string could have a strong impact on the performance of the algorithms.

In this section we present a high level model to process binary strings which exploits the block structure of text and  pattern to speed up the searching phase avoiding to work with bitwise operations. We suppose that the block size $k$ is fixed, so that all references to both text and pattern will only be to entire  blocks of $k$ bits. We refer to a $k$-bit block as a \emph{byte}, though larger values than $k = 8$ could be supported as well.
The idea to eliminate any reference to bits and to proceed block by block has been first suggested in~\cite{CKP85} for fast decoding of binary Huffman encoded texts. A similar approach  has been adopted also in~\cite{KBN2007,Fre2002}. %, where an efficient binary variant of the \bm algorithm was presented.
 For the sake of uniformity we use in the following, when it is possible, the same terminology adopted in~\cite{KBN2007}.

Let $T[i]$ and $P[i]$ denote, respectively, the $(i+1)$­th byte of the text and of the pattern, starting for $i = 0$ with both text and pattern aligned at the leftmost bit of the first byte. Since the lengths in bits of both text and pattern are not necessarily multiples  of $k$, the last byte may be only partially defined. In particular if the pattern has length $m$ then its last byte is that of position $\lceil m/k \rceil$ and only the leftmost $(m\mod k)$ bits of the last byte are defined. We suppose that the undefined bits of the last byte are set to $0$.

\begin{figure}[!t]
\begin{center}
\begin{scriptsize}
\begin{tabular}{l}
\begin{tabular}{r|c|c|c|c|}
~\textbf{(A)}~$\Patt$~~~&0&1&2&3\\
\hline
~~0~~ &~~\texttt{\cellcolor[gray]{0.8}\ul{11001011}}~~&~~\texttt{\cellcolor[gray]{0.8}\ul{00101100}}~~&~~\texttt{\ul{10110}000}~~&\\
~~1~~ &~~\texttt{0\ul{1100101}}~~&~~\texttt{\cellcolor[gray]{0.8}\ul{10010110}}~~&~~\texttt{\ul{010110}00}~~&\\
~~2~~ &~~\texttt{00\ul{110010}}~~&~~\texttt{\cellcolor[gray]{0.8}\ul{11001011}}~~&~~\texttt{\ul{0010110}0}~~&\\
~~3~~ &~~\texttt{000\ul{11001}}~~&~~\texttt{\cellcolor[gray]{0.8}\ul{01100101}}~~&~~\texttt{\cellcolor[gray]{0.8}\ul{10010110}}~~&\\
~~4~~ &~~\texttt{0000\ul{1100}}~~&~~\texttt{\cellcolor[gray]{0.8}\ul{10110010}}~~&~~\texttt{\cellcolor[gray]{0.8}\ul{11001011}}~~&~~\texttt{\ul{0}0000000}\\
~~5~~ &~~\texttt{00000\ul{110}}~~&~~\texttt{\cellcolor[gray]{0.8}\ul{01011001}}~~&~~\texttt{\cellcolor[gray]{0.8}\ul{01100101}}~~&~~\texttt{\ul{10}000000}\\
~~6~~ &~~\texttt{000000\ul{11}}~~&~~\texttt{\cellcolor[gray]{0.8}\ul{00101100}}~~&~~\texttt{\cellcolor[gray]{0.8}\ul{10110010}}~~&~~\texttt{\ul{110}00000}\\
~~7~~ &~~\texttt{0000000\ul{1}}~~&~~\texttt{\cellcolor[gray]{0.8}\ul{10010110}}~~&~~\texttt{\cellcolor[gray]{0.8}\ul{01011001}}~~&~~\texttt{\ul{0110}0000}\\
\hline
\end{tabular}~~~~~~
\begin{tabular}{|c|}
~\textbf{(C)}~$\Last$~\\
\hline
2\\
2\\
2\\
2\\
3\\
3\\
3\\
3\\
\hline
\end{tabular}\\
\\
\begin{tabular}{r|c|c|c|c|}
~\textbf{(B)}~$\Mask$~&0&1&2&3\\
\hline
~~0~~ &~~\cellcolor[gray]{0.8}\texttt{\ul{11111111}}~~&~~\cellcolor[gray]{0.8}\texttt{\ul{11111111}}~~&~~\texttt{\ul{11111}000}~~&\\
~~1~~ &~~\texttt{0\ul{1111111}}~~&~~\cellcolor[gray]{0.8}\texttt{\ul{11111111}}~~&~~\texttt{\ul{111111}00}~~&\\
~~2~~ &~~\texttt{00\ul{111111}}~~&~~\cellcolor[gray]{0.8}\texttt{\ul{11111111}}~~&~~\texttt{\ul{1111111}0}~~&\\
~~3~~ &~~\texttt{000\ul{11111}}~~&~~\cellcolor[gray]{0.8}\texttt{\ul{11111111}}~~&~~\cellcolor[gray]{0.8}\texttt{\ul{11111111}}~~&\\
~~4~~ &~~\texttt{0000\ul{1111}}~~&~~\cellcolor[gray]{0.8}\texttt{\ul{11111111}}~~&~~\cellcolor[gray]{0.8}\texttt{\ul{11111111}}~~&~~\texttt{\ul{1}0000000}\\
~~5~~ &~~\texttt{00000\ul{111}}~~&~~\cellcolor[gray]{0.8}\texttt{\ul{11111111}}~~&~~\cellcolor[gray]{0.8}\texttt{\ul{11111111}}~~&~~\texttt{\ul{11}000000}\\
~~6~~ &~~\texttt{000000\ul{11}}~~&~~\cellcolor[gray]{0.8}\texttt{\ul{11111111}}~~&~~\cellcolor[gray]{0.8}\texttt{\ul{11111111}}~~&~~\texttt{\ul{111}00000}\\
~~7~~ &~~\texttt{0000000\ul{1}}~~&~~\cellcolor[gray]{0.8}\texttt{\ul{11111111}}~~&~~\cellcolor[gray]{0.8}\texttt{\ul{11111111}}~~&~~\texttt{\ul{1111}0000}\\
\hline
\end{tabular}~~~~~~\\
\end{tabular}
\end{scriptsize}
\caption{Let $P=$\texttt{110010110010110010110}. \textbf{(A)} The matrix $\Patt$.
\textbf{(B)} The matrix $\Mask$.
\textbf{(C)} The array $\Last$. In $\Patt$ and $\Mask$ bits belonging to $P$ are underlined. Blocks containing a factor of $P$ are presented with light gray background color.}\label{example}
\end{center}
\end{figure}

In our high level model we define a sequence of several copies of the pattern memorized in the form of a matrix of bytes, $\Patt$, of size $k\times(\lceil m/k\rceil+1)$. Each row $i$ of the matrix $\Patt$ contains a copy of the pattern shifted by $i$ position to the right. The $i$ leftmost bits of the first byte remain undefined and are set to $0$. Similarly the rightmost $k-\left((m+i)\mod k\right)$ bits of the last byte are set to $0$. Formally the $j$-th bit of byte $\Patt[i,h]$ is defined by
$$
    \Patt[i,h]_j =
    \left\{
    \begin{array}{ll}
        p[kh-i+j]\      & \textrm{ if } 0\leq kh-i+j < m\\
        0\               & \textrm{ otherwise}
    \end{array}\right..
$$
for $0\leq i<k$ and $0\leq h < \lceil (m+i)/k\rceil$.

Observe that each factor of length $k$ of the pattern appears once in the table $\Patt$. In particular, the factor of length $k$ starting at position $j$ of $p$ is memorized in $\Patt[k-(j\mod k),\lceil j/k\rceil]$.

The high level model uses bytes in the matrix $\Patt$ to compare the pattern block by block against the text for any possible shift of the pattern. However when comparing the first or last byte of $P$ against its counterpart in the text, the bit positions not belonging to the pattern have to be neutralized. For this purpose we define a matrix of bytes, $\Mask$,  of size $k\times(\lceil m/k\rceil+1)$, containing binary masks of length $k$. In particular a bit in the mask $\Mask[i,h]$ is set to $1$ if and only if the corresponding bit of $\Patt[i, h]$ belongs to $P$. More formally
$$
    \Mask[i,h]_j =
    \left\{
    \begin{array}{ll}
        1\               & \textrm{ if } 0\leq kh-i+j < m\\
        0\               & \textrm{ otherwise}
    \end{array}\right..
$$
for $0\leq i<k$ and $0\leq h < \lceil (m+i)/k\rceil$.

Finally we need to compute an array, $\Last$, of size $k$ where $\Last[i]$ is defined to be the index of the last byte in the row $\Patt[i]$. Formally, for $0\leq i <k$ we define $\Last[i]=\lceil (m+i)/k\rceil$.

The procedure \ppp used to precompute the tables defined above is presented in Figure~\ref{naive}(A). It requires $\bigO(k \times \lceil m/k \rceil )=\bigO(m)$ time and $\bigO(m)$ extra-space. Figure~\ref{example} shows the precomputed tables defined above for a pattern $P=$\texttt{110010110010110010110} of length $m = 21$ and $k = 8$.

\begin{figure}[!t]
\begin{center}
\begin{scriptsize}
\begin{tabular}{|l|}
\hline
\begin{tabular}{l}
\\
~\textsc{\ppp}$(P, m)$\\[0.1cm]
~1   \quad $M \leftarrow 1^m0^{k-plast}$\\
~2   \quad \textbf{for} $i=0$ \textbf{to} k-1 \textbf{do}\\
~3   \quad \quad $\Last[i]=\lceil (m+i)/k \rceil-1$\\
~4   \quad \quad \textbf{for} $h=0$ \textbf{to} $\Last[i]$ \textbf{do}\\
~5   \quad \quad \quad $\Patt[i,h] \leftarrow (P[h]>>i)$\\
~6   \quad \quad \quad $\Mask[i,h] \leftarrow (M[h]>>i)$\\
~7   \quad \quad \quad \textbf{if} $h>0$ \textbf{then}\\
~8   \quad \quad \quad \quad $X \leftarrow \Patt[i,h]\ |\ (P[h-1]<<(k-i))$\\
~9   \quad \quad \quad \quad $\Patt[i,h] \leftarrow X$\\
10   \quad \quad \quad \quad $Y \leftarrow \Mask[i,h]\ |\ (M[h-1]<<(k-i))$\\
11   \quad \quad \quad \quad $\Mask[i,h] \leftarrow Y$\\
12   \quad \textbf{return} ($\Patt$, $\Last$, $\Mask$)\\
   \\
\hspace{5.8cm}\textbf{(A)}\\
%\hline
\end{tabular}
%\end{scriptsize}
%\caption{The \ppp procedure for the computation of the tables $\Patt$, $\Mask$ and $\Last$}
%\label{precompute}
%\end{center}
%\end{figure}
%
%\begin{figure}[!t]
%\begin{center}
%\begin{scriptsize}
\begin{tabular}{l}
%\hline
\\
~\textsc{\naive}$(P, m, T, n)$\\[0.1cm]
~1   \quad $(\Patt$, $L$, $M) \leftarrow$ \ppp($P$, $m$)\\
~2   \quad $s \leftarrow i \leftarrow w \leftarrow 0$\\
~3   \quad \textbf{while} $s < n$ \textbf{do}\\
~4   \quad \quad $j \leftarrow 0$\\
~5   \quad \quad \textbf{while} $j<L[i]$ and\\
~6   \quad \quad \quad $\Patt[i,j]= (T[w+j]\&M[i,j])$\\
~7   \quad \quad \quad \textbf{do} $j \leftarrow j+1$\\
~8   \quad \quad \textbf{if} $j=L[i]$ \textbf{then} Output($s$)\\
~9   \quad \quad $i \leftarrow i+1$\\
10   \quad \quad \textbf{if} $i=k$ \textbf{then}\\
11   \quad \quad \quad $w \leftarrow w+1$\\
12   \quad \quad \quad $i \leftarrow 0$\\
13   \quad \quad $s \leftarrow s+1$\\
\hspace{4.8cm}\textbf{(B)}\\
\end{tabular}\\
\hline
\end{tabular}
\end{scriptsize}
\caption{\textbf{(A)} The \ppp procedure for the computation of the tables $\Patt$, $\Mask$ and $\Last$. \textbf{(B)} The \naive algorithm for the binary string matching problem.}\label{naive}
\end{center}
\end{figure}

The model uses the precomputed tables to check whether $s$ is a valid shift without making use of bitwise operations but processing pattern and text byte by byte. In particular, for a given shift position $s$ (the pattern is aligned with the $s$-th bit of the text), we report a match if
\begin{equation}~\label{eq:compute}
    \Patt[i,h]= T[j+h]\ \&\ \Mask[i,h],\textrm{ for } h=0,1,...,\Last[i].
\end{equation}
where $j=\lfloor s/k \rfloor$ is the starting byte position in the text and $i=(s\mod k)$.

A simple \naive algorithm, obtained with this high level model, is shown in Figure~\ref{naive}(B). The algorithm starts by aligning the left ends of the pattern and text. Then, for each value of the shift $s=0,1,\ldots,n-m$, it checks whether $p$ occurs in $t$ by simply comparing each byte of the pattern with its  corresponding byte in the text, proceeding from left to right. At the end of the matching phase, the shift is advanced by one position to the right. In the worst­case,  the \naive algorithm requires $\bigO(\lceil m/k\rceil n)$ comparisons.

\section{New Efficient Binary String Matching Algorithms}\label{sec:newVariant}
In this section we present two new efficient algorithms for matching on binary strings based on the high level model presented above. The first algorithm is an adaptation of the \qh algorithm~\cite{Lec07} which is among the most efficient algorithms for the standard pattern matching problem. We show how the technique adopted by the algorithm can be naturally translated to allow for blocks of bits.

The second solution can be seen as an adaptation to binary string matching of the \sks algorithm~\cite{CLP98}. This algorithm can be efficiently adapted to completely avoid any reference to bits allowing to process pattern and text proceeding byte by byte.

\subsection{The \bh Algorithm}\label{sec:hash}
Algorithms in the \qh family for exact pattern matching have been introduced in~\cite{Lec07} where the author presented an adaptation of the Wu and Manber multiple string matching algorithm~\cite{WM94b} to single string matching problem.

The idea of the \qh algorithm is to consider factors of the pattern of length $q$. Each substring $w$ of such a length $q$ is hashed using a function $hash$ into integer values within $0$ and $255$. Then the algorithm computes in a preprocessing phase a function $\Hs: \{0,1,\ldots,255\} \rightarrow \{0,1,\ldots,m-q\}$, such that for each $0\leq c \leq 255$ the value $\Hs(c)$ is defined by
$$
\Hs(c) = \min \Big(\{0\leq k < m-q \:|\: hash(p[m-k-q \,..\, m-k-1]) = c\} \cup\{m-q\}\Big).
$$
The searching phase of the algorithm consists of reading, for each shift $s$ of the pattern in the text, the substring $w=t[s+m-q\,..\, s+m-1]$ of length $q$. If $\Hs(hash(w)) > 0$ then a shift of length $\Hs(hash(w))$ is applied. Otherwise, when $\Hs(hash(w)) = 0$ the pattern $p$ is naively checked in the text. In this case a shift of length $(m - 1 - i)$ is applied, where $i$ is the starting position of the rightmost occurrence in $p$ of a factor $p[j\,..\,j +q - 1]$ such that $hash(p[j\,..\,j +q - 1]) = hash(p[m -q +1\,..\,m- 1])$.

\smallskip

If the pattern $p$ is a binary string we can directly associate each substring of length $q$ with its numeric value in the range $[0,2^q-1]$ without making use of the $hash$ function. In order to exploit the block structure of the text we take into account substrings of length $q=k$. This means that, if $k=8$, each block $B$ of $k$ bits can be considered as a value $0\leq B\leq 255$. Thus we define a function $\Hs: \{0,1,\ldots,2^k-1\} \rightarrow \{0,1,\ldots,m\}$, such that for each byte $0\leq B < 2^k$
$$
\Hs(B) = \min \Big(\{0\leq u < m \:|\: p[m-u-k \,..\, m-u-1] \sqsupseteq B\} \cup\{m\}\Big) \enspace.
$$
Observe that if $B=p[m-k\,..\,m-1]$ then $\Hs[B]$ is defined to be $0$.

For example, in the case of the pattern $P=\texttt{110010110010110010110}$, presented in Figure~\ref{example}, we have $\Hs[$\texttt{01100101}$]=2$, $\Hs[$\texttt{11001011}$]=1$, and moreover $\Hs[$\texttt{10010110}$]=0$.

%\begin{figure}[!t]
%\begin{center}
%\begin{scriptsize}
%\begin{tabular}{|c|}
%\hline
%\begin{tabular}{rcccccccccc}
%&&&&&&&&&\\
%$j$        & $0$ & $1$ & $2$ & $3$ & $4$ & $5$ & $6$ &$7$&$8$&$9$\\
%$T:$       & \texttt{10110110} &\texttt{11001\ul{011}}&\texttt{\ul{10100}100}&\texttt{01010010}&\texttt{11\ul{100101}} &\texttt{\ul{10010110}} %&\texttt{0101\ul{1011}}&\texttt{\ul{0111}1011}&\texttt{11011110}&\texttt{00011011}\\
%&&&&&&&&\\
%(i)           & \texttt{11001011} &\texttt{00101100}& \texttt{10110~~~} &&&&&&&\\
%(ii) & $\rightarrow 21$ &&\texttt{~~~~~110} &\texttt{01011001}& \texttt{01100101}&\texttt{10~~~~~~~}&&&&\\
%(iii)& $\rightarrow 6$& & &\texttt{~~~11001}& \texttt{01100101}&\texttt{10010110~}&&&&\\
%(iv) & $\rightarrow 6$& & &  &\texttt{~1100101}& \texttt{10010110}&\texttt{010110~~~}&&&\\
%(v)  & $\rightarrow 6$& & &  &\texttt{~~~~~~~1}&\texttt{10010110}& \texttt{01011001}&\texttt{0110~~~~~}&&\\
%(vi)  & $\rightarrow 18$& & &  &&& &\texttt{~1100101}& \texttt{10010110}& \texttt{010110~~}\\
%&&&&&&&&&\\
%\end{tabular}\\
%
%\hline
%\end{tabular}
%\end{scriptsize}
%\caption{Execution of the \bh algorithm for the pattern $P=$\texttt{110010110010110010110} of length $21$ and a text $T$ of length $80$. Bytes %involved in the computation of the shift values have been underlined.}
%\label{example2}
%\end{center}
%\end{figure}

\begin{figure}[!t]
\begin{center}
\begin{scriptsize}
\begin{tabular}{|l|}
\hline
\begin{tabular}{ll}
&\\
~~\textsc{Compute-Hash}($\Patt$, $\Last$, $\Mask$, $m$)&\\[0.1cm]
~~~1.   \quad \textbf{for} $B \leftarrow 0$ \textbf{to} $2^k-1$ \textbf{do}&\\
~~~1.   \quad \quad $\Hs[B] \leftarrow m$&\\
~~~2.   \quad \textbf{for} $i \leftarrow k-1$ \textbf{downto} $1$ \textbf{do}&\\
~~~3.   \quad \quad \textbf{for} $B \leftarrow 0$ \textbf{to} $2^k-1$ \textbf{do}&\\
~~~4.   \quad \quad \quad \textbf{if} $\Patt[i,0]= B\ \&\ \Mask[i,0]$&\\
~~~5.   \quad \quad \quad \quad \textbf{then} $\Hs[B] \leftarrow m-k+i$&\\
~~~6.   \quad $i \leftarrow h \leftarrow 0$&\\
~~~7.   \quad \textbf{for} $j \leftarrow 0$ \textbf{to} $m-k-1$ \textbf{do}&\\
~~~8.   \quad \quad $\Hs[\Patt[i,h]] \leftarrow m-k-j$&\\
~~~9.   \quad \quad $i \leftarrow i-1$&\\
~~10.   \quad \quad \textbf{if} $i<0$ \textbf{then}&\\
~~11.   \quad \quad \quad $i \leftarrow k-1$&\\
~~12.   \quad \quad \quad $h \leftarrow h+1$&\\
~~13.   \quad \textbf{return} $\Hs$&\\
&\\
&\\
&\\
&\\
&\\
&\\
&\\
&\\
%\hline
\end{tabular}
\begin{tabular}{ll}
%\hline
&\\
~~\textsc{\bh}$(P, m, T, n)$&\\[0.1cm]
~~~1.   \quad $(\Patt$, $\Last$, $\Mask) \leftarrow$ \ppp($P$, $m$)&\\
~~~2.   \quad $\Hs \leftarrow$ \textsc{Compute-Hash}($\Patt$, $\Last$, $\Mask$, $m$)&\\
~~~3.   \quad $gap \leftarrow k-(m \mod k)$&\\
~~~4.   \quad $B \leftarrow \Patt[i][\Last[i]]$&\\
~~~5.   \quad $\shift \leftarrow \Hs[B]$, $\Hs[B] \leftarrow 0$&\\
~~~6.   \quad $j \leftarrow 0$, $s\ell \leftarrow m-1$&\\
~~~7.   \quad \textbf{while} $j < \lceil n/k\rceil$ \textbf{do}&\\
~~~8.   \quad \quad \textbf{while} $s\ell \geq k$ \textbf{do}&\\
~~~9.   \quad \quad \quad $s\ell \leftarrow s\ell - k$&\\
~~10.   \quad \quad \quad $j \leftarrow j+1$&\\
~~11.   \quad \quad $B \leftarrow T[j] \gg k-s\ell$&\\
~~12.   \quad \quad $B \leftarrow B\ |\ \left(T[j-1] \ll (s\ell+1)\right)$&\\
~~13.   \quad \quad \textbf{if} $\Hs[B]=0$ \textbf{then}&\\
~~14.   \quad \quad \quad $i \leftarrow (s\ell+gap) \mod k$&\\
~~15.   \quad \quad \quad $h \leftarrow \Last[i]$, $q\leftarrow 0$&\\
~~16.   \quad \quad \quad \textbf{while} $h>0$ and&\\
~~~~~   \quad \quad \quad \quad $Patt[i,h]= (T[j-q]\ \&\ \Mask[i,h])$&\\
~~17.   \quad \quad \quad \quad \textbf{do} $h \leftarrow h-1$, $q\leftarrow q+1$&\\
~~18.   \quad \quad \quad \textbf{if} $h<0$ \textbf{then} Output($j\times k + s\ell$)&\\
~~19.   \quad \quad \quad $s\ell \leftarrow s\ell+\shift$&\\
~~20.   \quad \quad \textbf{else} $s\ell \leftarrow s\ell+\Hs[B]$&\\
&\\
\end{tabular}\\
\hline
\end{tabular}
\end{scriptsize}
\caption{The \bh algorithm for the binary string matching problem.}
\label{algo:bh}
\end{center}
\end{figure}

The code of the \bh algorithm and its preprocessing phase are presented in Figure~\ref{algo:bh}.%, while Figure~\ref{example2} shows an example of its execution for the pattern $P$ of length $21$ shown in Figure~\ref{example} and a text $T$ of length $80$.

The preprocessing phase of the algorithm consists in computing the function $\Hs$ defined above and requires $\bigO(m+k2^{k+1})$ time complexity and $\bigO(m+2^k)$ extra space. During the search phase the algorithm reads, for each shift position $s$ of the pattern in the text, the block $B=t[s+m-q\,..\, s+m-1]$ of $k$ bits (lines \texttt{11-12}). If $\Hs(B) > 0$ then a shift of length $\Hs(B)$ is applied (line \texttt{20}). Otherwise, when $\Hs(B) = 0$ the pattern $p$ is naively checked in the text block by block (lines \texttt{15-18}).

After the test an advancement of length $\shift$ is applied (line \texttt{19}) where
$$
\shift = \min \Big(\{0< u < m \:|\: p[m-u-k \,..\, m-u-1] \sqsupset p[m-k \,..\, m-1]\} \cup\{m\}\Big)
$$
Observe that if the block $B$ has its $s\ell$ rightmost bits in in the $j$-th block of $T$ and the $(k-s\ell)$ leftmost bits in the block  $T[j-1]$, then it is computed by performing the following bitwise operation
$$
    B = \Big( T[j]\gg (k-s\ell) \Big)\ |\ \Big( T[j-1]\ll (s\ell+1) \Big)
$$

The \bh algorithm has a $\bigO(\lceil m/k \rceil n)$ time complexity and requires $\bigO(m+2^k)$ extra space.

For blocks of length $k$ the size of the $\Hs$ table is $2^k$, which seems reasonable for $k = 8$ or even $16$. For greater values of $k$ it is possible to adapt the algorithm to choose the desired time/space tradeoff by introducing a new parameter $K\leq k$, representing the number of bits taken into account for computing the shift advancement. Roughly speaking, only the $K$ rightmost bits of the current window of the text are taken into account, reducing the total sizes of the tables to $2^K$ at the cost of sometimes shifting the the pattern less than could be done if the full length of a block had been considered.

\subsection{The \bsks Algorithm}\label{sec:skipsearch}
The \sks algorithm has been presented in~\cite{CLP98} by Charras, Lecroq and Pehoushek. The idea of the algorithm is straightforward.
Let $p$ be a pattern of length $m$ and $t$ a text of length $n$, both over a finite alphabet $\Sigma$. For each character $c$ of the alphabet, a bucket collects all the positions of that character in the pattern. When a character occurs $\ell$ times in the pattern,  there are $\ell$ corresponding positions in the bucket of that character.  Formally, for $c \in \Sigma$ the \sks algorithm computes the table $S[c]$ where

$$
S[c] = \{i\ |\  0 \leq i < m\ \wedge\ P[i] = c\}.
$$

It is possible to notice that when  the pattern is much shorter than the alphabet, many buckets are empty.  The main loop of the search phase consists in examining every $m$-th  text character, $t[j]$ (so there will be $n/m$ main iterations). For each character $t[j]$,  it uses each position in the bucket $S[t[j]]$ to obtain all possible starting  positions of $p$ in $t$. For each position the algorithm performs a comparison of $p$ with $t$, character by character, until there is a mismatch, or until an occurrence is found.

For each possible block $B$ of $k$ bits, a bucket collects all pairs $(i,h)$ in the table $\Patt$ such that $\Patt[i,h]=B$. When a block of bits occurs more times in the pattern,  there are different corresponding pairs in the bucket of that block.  Observe that for a pattern of length $m$ there are $m-k+1$ different blocks of length $k$ corresponding to the blocks $\Patt[i,h]$ such that $kh-i\geq 0$ and $k(h+1)-i-1<m$.

However, to take advantage of the block structure of the text, we can compute buckets only for blocks contained in the suffix of the pattern of length $m'=k\lfloor m/k \rfloor$. In such a way $m'$ is a multiple of $k$ and we could reduce to examine a block for each $m'/k$ blocks of the text.
%This corresponds to compute buckets for all blocks of length $k$ in the pattern starting at position $(m\mod k)$.

Formally, for $0\leq B < 2^k$

$$S_k[B] = \{(i,h) :(m\mod k) \leq kh-i \leq m-k\ \wedge\ \Patt[i,h]=B\}.$$

For example in the case of the pattern $P=\texttt{110010110010110010110}$ we have
 $S_k[\texttt{01011001}]=\{(7,2)\}$,
 $S_k[\texttt{01100101}]=\{(3,1),(5,2)\}$,
 $S_k[\texttt{11001011}]=\{(2,1)\}$,
 $S_k[\texttt{10010110}]=\{(1,1),(3,2)\}$ and
 $S_k[\texttt{10110010}]=\{(4,2),(6,2)\}$.

The \bsks algorithm is shown in Figure~\ref{algo:bsks}. Its preprocessing phase consists in computing the buckets for all possible blocks of $k$ bits. The space and time complexity of this preprocessing phase is $\bigO(m+2^k)$.  The main loop of the search phase consists in examining every $(m'/k)$­th  text block. For each block $T[j]$ examined in the main loop, the algorithm inspects each pair $(i,pos)$ in the bucket $S_k[T[j]]$ to obtain a possible alignment of the pattern against the text (line \texttt{6}). For each pair $(i,pos)$ the algorithm checks whether $p$ occurs in $t$ by comparing $\Patt[i,h]$ and $T[j-pos+h]$, for $h=0,\ldots,\Last[i]$ (lines \texttt{7-10}). The \bsks algorithm has a $\bigO(\lceil m/k \rceil n)$ quadratic worst case time complexity and requires $\bigO(m+ 2^k)$ extra space.

\begin{figure}[!t]
\begin{center}
\begin{scriptsize}
\begin{tabular}{|l|}
\hline
\begin{tabular}{ll}
&\\
~~\textsc{Precompute-Skip-Table}($\Patt$, $m$)&\\[0.1cm]
~~~1.   \quad \textbf{for} $b = 0$ \textbf{to} $2^k-1$ \textbf{do} $S_k[b] \leftarrow \emptyset$&\\
~~~2.   \quad $i \leftarrow h\leftarrow 0$&\\
~~~3.   \quad \textbf{for} $j = 0$ \textbf{to} $m-k$ \textbf{do}&\\
~~~4.   \quad \quad \textbf{if} $j\geq (m \mod k)$ \textbf{then}&\\
~~~5.   \quad \quad \quad $b \leftarrow \Patt[i,h]$&\\
~~~6.   \quad \quad \quad $S_k[b] = S_k[b] \cup \{(i,h)\}$&\\
~~~7.   \quad \quad $i \leftarrow i-1$&\\
~~~8.   \quad \quad \textbf{if} $i<0$ \textbf{then}&\\
~~~9.   \quad \quad \quad $i \leftarrow k-1$&\\
~~10.   \quad \quad \quad $h \leftarrow h+1$&\\
~~11.   \quad \textbf{return} $S_k$&\\
&\\
&\\
%\hline
\end{tabular}
\begin{tabular}{ll}
%\hline
&\\
~~\textsc{\bsks}$(P, m, T, n)$&\\[0.1cm]
~~~1.   \quad $(\Patt$, $\Last$, $\Mask) \leftarrow$ \textsc{Preprocess}($P$, $m$)&\\
~~~2.   \quad $S_k \leftarrow$ \textsc{Precompute-Skip-Table}($\Patt$, $m$)&\\
~~~3.   \quad $\shift \leftarrow \lfloor m / k \rfloor -1$&\\
~~~4.   \quad $j \leftarrow \shift-1$&\\
~~~5.   \quad \textbf{while} $j < \lceil n/k \rceil$ \textbf{do}&\\
~~~6.   \quad \quad \textbf{for each} $(i,pos)\ \in\ S_k[T[j]]$ \textbf{do}&\\
~~~7.   \quad \quad \quad $h \leftarrow 0$&\\
~~~8.   \quad \quad \quad \textbf{while} $h<\Last[i]$ and &\\
~~~~~   \quad \quad \quad \quad $P[i,h]= (T[j-pos+h]\ \&\ \Mask[i,h])$&\\
~~~9.   \quad \quad \quad \quad \textbf{do} $h \leftarrow h+1$&\\
~~10.   \quad \quad \quad \textbf{if} $h=\Last[i]$ \textbf{then} Output($j\times k +i$)&\\
~~11.   \quad \quad $j \leftarrow j+\shift$&\\
&\\
\end{tabular}\\
\hline
\end{tabular}
\end{scriptsize}
\caption{The \bsks algorithm for the binary string matching problem.}
\label{algo:bsks}
\end{center}
\end{figure}

In practice, if the block size is $k$, the \bsks algorithm requires a table of size $2^k$ to compute the function $S_k$. This is just $256$ for $k = 8$, but for $k = 16$ or even $32$, such a table might be too large. In particular for growing values of $k$, there will be many cache misses, with strong impact on the performance of the algorithm. Thus for values of $k$ greater than $8$ it may be suitable to compute the function on the fly, still using a table for single bytes. Suppose for example that $k = 32$ and suppose $B$ is a block of $k$ bits. Let $B_j$ be the $j$-th byte of $B$, with $j=1,\ldots,4$. The set of all possible pairs associated to the block $B$ can be computed as
$$
    S_k[B]\ =\ S_k[B_1]\ \cap\ S_k^{1}[B_2]\ \cap\ S_k^{2}[B_3]\ \cap\ S_k^{3}[B_4]
$$
where  we have set $S_k^{q}[B_j]=\{ (i,h)\ |\ (i,h+q)\ \in\ S_k[B_j]\}$.

If we suppose that the distribution of zeros and ones in the input string is like in a randomly generated one, then the probability of occurrence in the text of any binary string of length $k$ is $2^{-k}$. This is a reasonable assumption for compressed text~\cite{KBD89}. Then the expected cardinality of set $S_k[B]$, for a pattern $p$ of length $m$, is  $(m-7)\times 2^{-8}$, that is less than 2 if $m<500$. Thus in practical cases the set $S_k[B]$ can be computed in constant expected time.

\section{Experimental Results}\label{sec:exp}
Here we present experimental data which allow to compare, in terms of running time and number of text character inspections, the following string matching algorithms under various conditions: the \naive algorithm (\textsf{BNAIVE}) of Figure~\ref{naive},  the \bbm algorithm by Klein (\textsf{BBM}) presented in~\cite{KBN2007}, the \bh algorithm (\textsf{BHM}) of Figure~\ref{algo:bh}, and the  \bsks algorithm (\textsf{BSKS})  of Figure~\ref{algo:bsks}.

For the sake of completeness, for experimental results on running times we have also tested the following algorithms for standard pattern matching: the $q$-\textsc{Hash} algorithm~\cite{Lec07} with $q=8$ (\textsf{HASH8}) and the \textsc{Extended-BOM} algorithm~\cite{FL2008} (\textsf{EBOM}). These are among the most efficient in practical cases. The $q$-\textsc{Hash} and \textsc{Extended-BOM} algorithms have been tested on the same texts and patterns but in their standard form, i.e. each character is an ASCII value of $8$-bit, thus obtaining a comparison between methods on standard and binary strings.

To simulate the different conditions which can arise when processing binary data we have performed our tests on texts with a different distribution of zeros and ones.  For the case of compressed strings it is quite reasonable to assume a uniform distribution of characters. For compression scheme using Huffman coding, such randomness has been shown to hold in~\cite{KBD89}.   In contrast when processing binary images we aspect a non-uniform distribution of characters. For instance in a fax-image usually more than $90$\% of the total number of bits is set to zero.

All algorithms have been implemented in the \textbf{C} programming language and were used to search for the same binary strings in large fixed text buffers on a PC with Intel Core2 processor of 1.66GHz.  In particular, the algorithms have been tested on three \textsf{Rand}$(1/0)_{\gamma}$ problems, for $\gamma=50,70$ and $90$. Searching have been performed for binary patterns, of length $m$ from $20$ to $500$, which have been taken as substring of the text at random starting positions.

In particular each \textsf{Rand}$(1/0)_{\gamma}$ problem consists of searching a set of $1000$ random patterns of a given length in a random binary text of $4\times 10^6$ bits. The distribution of characters depends on the value of the parameter $\gamma$. In particular bit $0$ appears with a percentage equal to $\gamma$\%.

Moreover, for each test, the average number of character inspections has been computed by taking the total number of times a text byte is accessed (either to perform a comparison with the pattern, or to perform a shift) and dividing it by the length of the text buffer.

In the following tables, running times (on the left) are expressed in hundredths of seconds. Tables with the number of text character inspections (on the right) are presented in light-gray background color. Best results are bold faced.

\begin{center}
\begin{scriptsize}
\begin{tabular*}{0.56\textwidth}{@{\extracolsep{\fill}}|l|cccc|cc|}
\hline
$m$& \textsf{BNAIVE}&\textsf{BBM}&\textsf{BSKS}&\textsf{BHM}&\textsf{HASH8}&\textsf{EBOM}\\
\hline
\texttt{20}  &~ 41.53 ~&~ 13.53 ~&~ 3.66 ~&~ \tb{3.40} ~&~ 5.12 ~&~ 8.89 ~\\
\texttt{60}  &~ 41.72 ~&~ 7.77 ~&~ \tb{1.16} ~&~ 1.60  ~&~ 1.72 ~&~ 3.85 ~\\
\texttt{100} &~ 41.68 ~&~ 6.80 ~&~ \tb{0.70} ~&~ 1.44 ~&~ 1.64 ~&~ 3.06 ~\\
\texttt{140} &~ 42.11 ~&~ 6.21 ~&~ \tb{0.89} ~&~ 1.24 ~&~ 1.54 ~&~ 2.67 ~\\
\texttt{180} &~ 41.95 ~&~ 5.76 ~&~ \tb{0.66} ~&~ 1.10 ~&~ 1.80 ~&~ 2.25 ~\\
\texttt{220} &~ 41.93 ~&~ 5.36 ~&~ \tb{0.74} ~&~ 1.24 ~&~ 1.79 ~&~ 1.87 ~\\
\texttt{260} &~ 41.95 ~&~ 5.08 ~&~ \tb{0.54} ~&~ 1.05 ~&~ 1.47 ~&~ 2.09 ~\\
\texttt{300} &~ 41.74 ~&~ 5.07 ~&~ \tb{0.54} ~&~ 1.11 ~&~ 1.82 ~&~ 1.48 ~\\
\texttt{340} &~ 41.93 ~&~ 4.86 ~&~ \tb{0.39} ~&~ 1.07 ~&~ 1.56 ~&~ 1.56 ~\\
\texttt{380} &~ 41.97 ~&~ 4.59 ~&~ \tb{0.46} ~&~ 0.97 ~&~ 1.87 ~&~ 1.43 ~\\
\texttt{420} &~ 42.07 ~&~ 4.52 ~&~ \tb{0.31} ~&~ 1.23 ~&~ 1.59 ~&~ 1.23 ~\\
\texttt{460} &~ 41.99 ~&~ 4.68 ~&~ \tb{0.23} ~&~ 1.04 ~&~ 1.52 ~&~ 1.19 ~\\
\texttt{500} &~ 42.06 ~&~ 4.61 ~&~ \tb{0.37} ~&~ 0.81 ~&~ 1.53 ~&~ 1.32 ~\\
\hline
\end{tabular*}~~~
\begin{tabular}{c}
\rowcolor[gray]{0.8}
\begin{tabular*}{0.38\textwidth}{@{\extracolsep{\fill}}|ccccc|}
\hline
&\textsf{BNAIVE}&\textsf{BBM}&\textsf{BSKS}&\textsf{BHM}\\
\hline
&~9.00 ~&~ 1.82 ~&~ 1.04 ~&~ \tb{0.90} ~\\
&~9.00 ~&~ 0.85 ~&~ \tb{0.20} ~&~ 0.31 ~\\
&~9.00 ~&~ 0.63 ~&~ \tb{0.13} ~&~ 0.20 ~\\
&~9.00 ~&~ 0.54 ~&~ \tb{0.10} ~&~ 0.15 ~\\
&~9.00 ~&~ 0.47 ~&~ \tb{0.08} ~&~ 0.13 ~\\
&~9.00 ~&~ 0.44 ~&~ \tb{0.07} ~&~ 0.11 ~\\
&~9.00 ~&~ 0.41 ~&~ \tb{0.07} ~&~ 0.10 ~\\
&~9.00 ~&~ 0.39 ~&~ \tb{0.06} ~&~ 0.09 ~\\
&~9.00 ~&~ 0.38 ~&~ \tb{0.06} ~&~ 0.09 ~\\
&~9.00 ~&~ 0.37 ~&~ \tb{0.06} ~&~ 0.08 ~\\
&~9.00 ~&~ 0.36 ~&~ \tb{0.05} ~&~ 0.08 ~\\
&~9.00 ~&~ 0.35 ~&~ \tb{0.05} ~&~ 0.08 ~\\
&~9.00 ~&~ 0.35 ~&~ \tb{0.05} ~&~ 0.07 ~\\
\hline
\end{tabular*}
\end{tabular}\\[0.1cm]
Experimental results for a \textsf{Rand}$(0/1)_{50}$ problem
\end{scriptsize}
\end{center}

\begin{center}
\begin{scriptsize}
\begin{tabular*}{0.56\textwidth}{@{\extracolsep{\fill}}|l|cccc|cc|}
\hline
$m$& \textsf{BNAIVE}&\textsf{BBM}&\textsf{BSKS}&\textsf{BHM}&\textsf{HASH8}&\textsf{EBOM}\\
\hline
\texttt{20} &~ 43.26 ~&~ 17.25 ~&~ \tb{4.01} ~&~ 4.21 ~&~ 4.86 ~&~ 10.92 ~\\
\texttt{60} &~ 43.15 ~&~ 10.26 ~&~ \tb{1.66} ~&~ 2.09 ~&~ 2.03 ~&~ 4.27 ~\\
\texttt{100} &~ 43.80 ~&~ 8.44 ~&~ \tb{1.60} ~&~ 2.26 ~&~ 1.95 ~&~ 2.54 ~\\
\texttt{140} &~ 43.70 ~&~ 8.13 ~&~ \tb{1.28} ~&~ 1.61 ~&~ 1.52 ~&~ 2.68 ~\\
\texttt{180} &~ 43.22 ~&~ 7.37 ~&~ \tb{1.02} ~&~ 1.67 ~&~ 2.08 ~&~ 2.33 ~\\
\texttt{220} &~ 43.29 ~&~ 6.82 ~&~ \tb{1.08} ~&~ 1.34 ~&~ 1.94 ~&~ 2.50 ~\\
\texttt{260} &~ 42.93 ~&~ 6.67 ~&~ \tb{1.07} ~&~ 1.53 ~&~ 1.79 ~&~ 1.94 ~\\
\texttt{300} &~ 43.66 ~&~ 6.46 ~&~ \tb{0.89} ~&~ 1.22 ~&~ 1.59 ~&~ 1.94 ~\\
\texttt{340} &~ 43.53 ~&~ 6.35 ~&~ \tb{0.97} ~&~ 1.23 ~&~ 1.28 ~&~ 1.86 ~\\
\texttt{380} &~ 43.76 ~&~ 6.15 ~&~ \tb{0.70} ~&~ 1.42 ~&~ 1.31 ~&~ 1.65 ~\\
\texttt{420} &~ 43.29 ~&~ 6.03 ~&~ \tb{0.85} ~&~ 1.34 ~&~ 1.67 ~&~ 1.48 ~\\
\texttt{460} &~ 43.45 ~&~ 6.00 ~&~ \tb{0.92} ~&~ 1.27 ~&~ 1.37 ~&~ 1.43 ~\\
\texttt{500} &~ 43.31 ~&~ 6.00 ~&~ \tb{0.70} ~&~ 1.28 ~&~ 1.41 ~&~ 1.48 ~\\
\hline
\end{tabular*}~~~
\begin{tabular}{c}
\rowcolor[gray]{0.8}
\begin{tabular*}{0.38\textwidth}{@{\extracolsep{\fill}}|ccccc|}
\hline
&\textsf{BNAIVE}&\textsf{BBM}&\textsf{BSKS}&\textsf{BHM}\\
\hline
& ~9.41~ & ~2.27~ & ~1.12~ & ~\tb{1.01}~ \\
& ~9.40~ & ~1.14~ & ~\tb{0.29}~ & ~0.38~ \\
& ~9.38~ & ~0.89~ & ~\tb{0.21}~ & ~0.26~ \\
& ~9.38~ & ~0.77~ & ~\tb{0.18}~ & ~0.21~ \\
& ~9.37~ & ~0.71~ & ~\tb{0.17}~ & ~0.18~ \\
& ~9.39~ & ~0.65~ & ~\tb{0.16}~ & ~\tb{0.16}~ \\
& ~9.38~ & ~0.61~ & ~\tb{0.15}~ & ~\tb{0.15}~ \\
& ~9.39~ & ~0.59~ & ~0.15~ & ~\tb{0.14}~ \\
& ~9.39~ & ~0.57~ & ~0.15~ & ~\tb{0.13}~ \\
& ~9.38~ & ~0.55~ & ~0.14~ & ~\tb{0.12}~ \\
& ~9.38~ & ~0.54~ & ~0.14~ & ~\tb{0.12}~ \\
& ~9.38~ & ~0.53~ & ~0.14~ & ~\tb{0.11}~ \\
& ~9.37~ & ~0.51~ & ~0.14~ & ~\tb{0.11}~ \\
\hline
\end{tabular*}
\end{tabular}\\[0.1cm]
Experimental results for a \textsf{Rand}$(0/1)_{70}$ problem
\end{scriptsize}
\end{center}

\begin{center}
\begin{scriptsize}
\begin{tabular*}{0.56\textwidth}{@{\extracolsep{\fill}}|l|cccc|cc|}
\hline
$m$& \textsf{BNAIVE}&\textsf{BBM}&\textsf{BSKS}&\textsf{BHM}&\textsf{HASH8}&\textsf{EBOM}\\
\hline
\texttt{20} & ~50.61~ & ~41.19~ & ~\tb{18.51}~ & ~21.00~ & ~24.30~ & ~24.95~ \\
\texttt{60} & ~51.68~ & ~30.62~ & ~\tb{13.61}~ & ~14.32~ & ~13.65~ & ~7.23~ \\
\texttt{100} & ~53.00~ & ~28.44~ & ~\tb{12.22}~ & ~12.64~ & ~11.64~ & ~5.15~ \\
\texttt{140} & ~51.78~ & ~27.16~ & ~11.86~ & ~\tb{11.44}~ & ~10.31~ & ~4.09~ \\
\texttt{180} & ~51.51~ & ~24.78~ & ~11.80~ & ~\tb{10.21}~ & ~9.83~ & ~3.21~ \\
\texttt{220} & ~52.54~ & ~24.60~ & ~11.50~ & ~\tb{9.63}~ & ~9.13~ & ~3.12~ \\
\texttt{260} & ~52.38~ & ~23.59~ & ~11.85~ & ~\tb{8.74}~ & ~8.61~ & ~2.31~ \\
\texttt{300} & ~52.00~ & ~22.68~ & ~11.15~ & ~\tb{8.73}~ & ~8.11~ & ~2.63~ \\
\texttt{340} & ~51.98~ & ~21.72~ & ~11.30~ & ~\tb{8.02}~ & ~7.24~ & ~2.29~ \\
\texttt{380} & ~52.33~ & ~21.79~ & ~11.39~ & ~\tb{7.66}~ & ~7.57~ & ~2.17~ \\
\texttt{420} & ~52.35~ & ~21.16~ & ~10.94~ & ~\tb{7.58}~ & ~7.43 ~& ~1.82~ \\
\texttt{460} & ~52.29~ & ~20.54~ & ~11.09~ & ~\tb{6.75}~ &~ 6.45~ & ~2.12~ \\
\texttt{500} & ~51.68~ & ~20.68~ & ~11.12~ & ~\tb{7.42}~ & ~6.99~ & ~1.79~ \\
\hline
\end{tabular*}~~~
\begin{tabular}{c}
\rowcolor[gray]{0.8}
\begin{tabular*}{0.38\textwidth}{@{\extracolsep{\fill}}|ccccc|}
\hline
&\textsf{BNAIVE}&\textsf{BBM}&\textsf{BSKS}&\textsf{BHM}\\
\hline
& ~12.46 ~&~ 6.88 ~&~ \tb{3.79} ~&~ 4.87 \\
& ~12.53 ~&~ 5.14 ~&~ \tb{2.82} ~&~ 3.28 \\
& ~12.72 ~&~ 4.70 ~&~ 2.78 ~&~ \tb{2.76} \\
& ~12.46 ~&~ 4.47 ~&~ 2.63 ~&~ \tb{2.53} \\
& ~12.45 ~&~ 4.11 ~&~ 2.59 ~&~ \tb{2.22} \\
& ~12.69 ~&~ 4.02 ~&~ 2.65 ~&~ \tb{2.09} \\
& ~12.55 ~&~ 3.87 ~&~ 2.58 ~&~ \tb{1.97} \\
& ~12.59 ~&~ 3.67 ~&~ 2.64 ~&~ \tb{1.80} \\
& ~12.53 ~&~ 3.53 ~&~ 2.60 ~&~ \tb{1.70} \\
& ~12.56 ~&~ 3.53 ~&~ 2.64 ~&~ \tb{1.69} \\
& ~12.60 ~&~ 3.46 ~&~ 2.56 ~&~ \tb{1.60} \\
& ~12.51 ~&~ 3.28 ~&~ 2.59 ~&~ \tb{1.48} \\
& ~12.34 ~&~ 3.39 ~&~ 2.57 ~&~ \tb{1.55} \\
\hline
\end{tabular*}
\end{tabular}\\[0.1cm]
Experimental results for a \textsf{Rand}$(0/1)_{90}$ problem
\end{scriptsize}
\end{center}

Experimental results show that the \bsks and the \bh algorithms obtain the best run-time performance in all cases. In particular it turns out that the \bsks algorithm is the best choice when the distribution of character is uniform. In this case the algorithm is $10$ times faster than \bbm, and $100$ times faster than \naive. Moreover performs less than $50$\% of inspections performed by the \bbm algorithm, especially for long patterns.

For non-uniform distribution of characters the \bh algorithm obtains the best results in terms of both running time and number of character inspections. It turns out to be at least two times faster than $\bbm$ algorithm and to perform a number of text character inspections which is less than $50$\% of that performed by the \bbm algorithm.

\section{Conclusion}\label{sec:conclusion}
Efficient variants of the \qh and \sks pattern matching algorithms have been presented for the case in which both text and pattern are over a binary alphabet. The algorithm exploit the block structure of the binary strings and process text and pattern with no use of any bit manipulations. Both  algorithms have a $\bigO(n/m)$ time complexity. However, from our experimental results it turns out that the presented algorithms are the most effective in practical cases.

\bibliographystyle{alpha}
\bibliography{biblio}

\end{document}